\documentclass{article}
\usepackage[]{fontenc}
\usepackage{a4wide}

\makeatletter

\newcommand{\LyX}{L\kern-.1667em\lower.25em\hbox{Y}\kern-.125emX\spacefactor1000}

\makeatother

\begin{document}

\begin{center} 
{\LARGE {\bf Vertex operators, semiclassical limit for soliton S-matrices and the number of bound states in}}\vspace{0.25cm} {\LARGE{\bf  Affine Toda Field Theories} }
  
\vspace{0.9cm}

 {\large Marco A.C. Kneipp} \footnote{
e-mail: {\tt kneipp@cbpf.br}}  
\end{center}

%\maketitle
\begin{center}{\em Departamento de Campos e Part\'\i culas, \\ 
Centro Brasileiro de Pesquisas F\' \i sicas (CBPF), \\
Rua Dr. Xavier Sigaud,150 \\
22290-180 Rio de Janeiro, Brazil }

\end{center}

\vspace{0.5cm}

\begin{abstract}
Soliton time delays and the semiclassical limit for soliton S-matrices are calculated
for non-simply laced Affine Toda Field Theories. The phase shift is written
as a sum over bilinears on the soliton conserved charges. The results apply
to any two solitons of any Affine Toda Field Theory. As a by-product, a general
expression for the number of bound states and the values of the coupling in
which the S-matrix can be diagonal are obtained. In order to arrive at these
results, a vertex operator is constructed, in the principal gradation, for non-simply
laced affine Lie algebras, extending the previous constructions for simply laced
and twisted affine Lie algebras.
\end{abstract}

\section{Introduction.}

In 1+1 dimensions, a well-known class of integrable theories are the Affine
Toda Field Theories (ATFTs). For each affine Lie algebra \( \widehat{g} \),
we can associated an ATFT. For simplicity just the untwisted algebras will be
considered.(The twisted cases were consided in \cite{KO2}.) If the coupling
\( \beta  \) is imaginary, there exist degenerate vacua and solitons interpolating
these vacua. An \( N \)-soliton solution can be written as \cite{OTU1} 
\begin{equation}
\label{1.6}
e^{-\beta \lambda _{j}\cdot \phi }=\frac{<\Lambda _{j}|g(t)|\Lambda _{j}>}{<\Lambda _{0}|g(t)|\Lambda _{0}>^{m_{j}}}
\end{equation}
with
\begin{eqnarray}
g(t) & = & \prod _{k=1}^{N}e^{Q_{i(k)}W_{i(k)}\widehat{F}^{i(k)}(\theta )}\, ,\label{1.6a} \\
W_{i(k)}(x,t) & = & e^{\mu _{i(k)}\cosh \theta _{k}(x-v_{k}t)}\, ,\label{1.6b} \\
Q_{i(k)} & = & e^{i\psi _{k}}e^{-x_{k}^{0}\mu _{i(k)}\cosh \theta _{k}}\, ,\label{1.6c} 
\end{eqnarray}
where \( \theta _{k} \) is the rapidity of the \( k^{\mbox {th}} \) soliton,
\( x_{k}^{0} \) its position at \( t=0 \), \( \psi _{k} \) is a phase and
\( \mu _{i(k)} \)is the mass particle of species \( i(k) \). Let \( g \)
be the underlying Lie algebra. For each dot of the Dynkin diagram of \( g \)
a soliton species can be associated. Each species may have various different
topological charges. For example, sine-Gordon, which is associated with \( \widehat{g}=a^{(1)}_{1} \),
has one soliton species with two topological charges: the well-known soliton
and anti-soliton solutions. The mass for each species is\cite{OTU1}
\begin{equation}
\label{1.6d}
M_{i}=\frac{4h\mu _{i}}{\alpha ^{2}_{i}|\beta ^{2}|}\, ,
\end{equation}
where \( h \) is the Coxeter number of \( g \). From (\ref{1.6}), we can
equivalently write the \( N \)-soliton solution as
\[
\phi =-\frac{1}{\beta }\sum _{j=0}^{r}\alpha ^{\mbox {v}}_{j}\ln <\Lambda _{j}|g(t)|\Lambda _{j}>\]
 where \( \alpha _{0} \) is the negative of the hightest root. Written in this
form, it becomes clear that \( <\Lambda _{j}|g(t)|\Lambda _{j}> \) are the
\( \tau  \)-functions which appear in the Hirota method, used in\cite{Hol92}-\cite{KG93}. 

Quite a lot of work has been done in order to obtain the S-matrix for the particles\cite{AFZ79}-\cite{FKS99}
and for the solitons\cite{KTTW77}-\cite{PJ96} of ATFT. As usual, the proposed
S-matrix must be constructed in agreement with the S-matrix axioms. However,
in order to confirm that the proposed S-matrix is associated with a given theory,
one must check if it has the correct semiclassical limit.

The highest non-vanishing power of \( \widehat{F}^{i}(\theta ) \) has a vertex
operator construction in the principal gradation which was first proven for
level one representations \cite{LW78}\cite{KKLW81}\cite{OTU2} and then extended
for any representation of simply laced \cite{KO1} and twisted\cite{KO2} affine
Lie algebras. From this construction, it was shown\cite{FJKO} that any given
soliton regains its original shape, after having collisions with other solitons,
as expected by a soliton solution. The only effect is a time delay which was
calculated. From the time delay, the semiclassical limit for the transmission
amplitude of the simply laced soliton S-matrix was obtained\cite{FJKO} . 

In the present paper, these results are extended for the remaining case of non-simply
laced affine Lie algebras obtaining a semiclassical expression which holds for
any affine Lie algebra. The paper is divided as follows: in section 2, after
a brief review, it is shown that the non-simply laced case, the last non-vanishing
power of \( \widehat{F}^{i}(\theta ) \) also has a vertex construction. Then,
the asymptotic behavior of the solitons is analyzed in section 3 and the time
delay resulting from the collision of two or more solitons is calculated in
section 4. From the time delay, the semiclassical limit of the soliton S-matrix
which holds for any ATFT is calculated, in section 5. The phase shift is written
in terms of the soliton conserved charges. In section 6 a general expression
for the number of bound states in the direct and cross channel of the S-matrix
is obtained along with the values of the coupling in which the S-matrix can
be purely elastic. Our conclusions are stated in section 7.

\section{Vertex operator construction for non-simply laced affine Lie algebras.}

In order to deal with soliton solitons in ATFT, it is convenient to consider
an alternative basis for an affine Lie algebra \( \widehat{g} \) (for simplicity,
let us consider \( \widehat{g} \) untwisted) with generators \( \widehat{E}_{M} \),
\( M \) being an exponent of \( \widehat{g} \)(i.e \( M=mh+\mu  \) where
\( \mu  \) are exponents of \( g \) and \( m\in Z \)) , \( \widehat{F}_{N}^{i} \)
, \( N\in Z \) and the level \( x \). Using the conventions of \cite{OTU1}-\cite{FJKO},
they satisfy the commutation relations
\begin{eqnarray}
\left[ \widehat{E}_{M},\widehat{E}_{N}\right]  & = & Mx\delta _{M+N,0}\, ,\label{1.1} \\
\left[ \widehat{E}_{M},\widehat{F}_{N}^{i}\right]  & = & \gamma _{i}\cdot q([M])\widehat{F}^{i}_{M+N}\, .\label{1.2} 
\end{eqnarray}
where \( [M] \) means \( M \) mod \( h \), \( q([M]) \) is the eigenvector
of the Coxeter element associated with the eigenvalue \( \exp 2\pi i[M]/h \)
and \( \gamma _{i}=c_{i}\alpha _{i} \) with \( c_{i}=\pm 1 \) depending on
the ``color'' of \( \alpha _{i} \). The generators which ad-diagonalize the
principal Heisenberg subalgebra are
\begin{equation}
\label{1.3}
\widehat{F}^{i}(z)=a_{i}\sum ^{\infty }_{N=-\infty }z^{-N}\widehat{F}^{i}_{N}\, ,
\end{equation}
 \( z \) being a complex variable. The numbers \( a_{i} \) are normalization
constants which can be fixed by imposing \( <\Lambda _{0}|\widehat{F}^{i}(z)|\Lambda _{0}>=1 \)
, implying that \cite{KO2}
\begin{equation}
\label{1.4}
<\Lambda _{j}|\widehat{F}^{i}(z)|\Lambda _{j}>=e^{-2\pi i\lambda _{i}\cdot \lambda _{j}}\, .
\end{equation}
From the commutation relation (\ref{1.2}), it follows directly that
\begin{equation}
\label{1.5}
\left[ \widehat{E}_{M},\widehat{F}^{i}(z)\right] =\gamma _{i}\cdot q([M])z^{M}\widehat{F}^{i}(z)\, .
\end{equation}
For a \emph{simply-laced} \( \widehat{g} \), in the representation with highest
weight \( \Lambda _{j} \) of level \( m_{j} \), the highest nonvanishing power
of \( \widehat{F}^{i}(z) \) has a vertex operator construction\cite{OTU2}\cite{KO1}
\begin{equation}
\label{1.5a}
\widehat{V}^{i}(z)\equiv \frac{\widehat{F}^{i}(z)^{m_{j}}}{m_{j}!}=e^{-2\pi i\lambda _{i}\cdot \lambda _{j}}Y^{i}_{-}Y^{i}_{+}
\end{equation}
where

\[
Y^{i}_{\pm }=\exp \left( \sum _{M\in \cal E}\frac{\gamma _{i}\cdot q([\pm M])z^{\mp M}}{\mp M}E_{\pm M}\right) \, .\]
with the sun running over the positive exponents \( \cal E \) of \( \widehat{g} \).
The normal ordering of two vertex operators is \cite{OTU2}\cite{KO1}
\begin{equation}
\label{1.5c}
\widehat{V}^{i}(z)\widehat{V}^{k}(w)=X_{ik}(z,w)^{m_{j}}:\widehat{V}^{i}(z)\widehat{V}^{k}(w):
\end{equation}
where
\[
:\widehat{V}^{i}(z)\widehat{V}^{k}(w):=e^{-2\pi i\left( \lambda _{i}+\lambda _{k}\right) \cdot \lambda _{j}}Y^{i}_{-}Y^{k}_{-}Y^{i}_{+}Y^{k}_{+}\]

\begin{equation}
\label{1.5d}
X_{<i><k>}(z_{i},z_{k})=\prod _{p=1}^{h}\left( 1-\omega ^{p}\frac{z_{k}}{z_{i}}\right) ^{\sigma ^{-p}\left( \gamma _{<i>}\right) \cdot \gamma _{<k>}}\, .
\end{equation}
 Let us now prove for non-simply laced affine Lie algebras, the highest non-vanishing
power of \( \widehat{F}^{i}(z) \) also admits a vertex operator construction.
As is well-known \cite{OT1}, all non-simply laced Lie algebras can be obtained
from a simply-laced Lie algebra \( g \) as a fixed subalgebra \( g_{\tau } \)
under an outer automorphism \( \tau  \) of \( g \). The Dynkin diagram of
\( g_{\tau } \), \( \Delta (g_{\tau }) \), is obtained by identifying the
vertices on each separate orbit of \( \tau  \). The set of vertices in the
orbit containing the vertex \( i \) will be denoted by \( <i> \) and \( p_{i} \)
stands for the number of vertices. Denoting by \( \alpha _{i} \) and \( \lambda _{i} \),
the simple roots and fundamental weights of \( g \), the \( g_{\tau } \) simple
roots are\cite{GNOS86}

\begin{equation}
\label{2.10a}
\alpha _{<i>}=\frac{1}{p_{i}}\sum _{i\in <i>}\alpha _{i}
\end{equation}
and the \( g_{\tau } \) fundamental coweights 
\begin{equation}
\label{2.10b}
\lambda _{<i>}^{\mbox {v}}=\sum _{i\in <i>}\lambda _{i}
\end{equation}
have the correct inner products with the simple roots. From (\ref{2.10a}),
it follows that
\begin{equation}
\label{2.10c}
\frac{2}{\alpha ^{2}_{<i>}}=p_{i}\, \, \, \Rightarrow \, \, \, \alpha _{<i>}^{\mbox {v}}=\sum _{i\in <i>}\alpha _{i}\, .
\end{equation}

Turning to the affine Lie algebras, one can once more obtain all the non-simply
laced \( \widehat{g}_{\tau } \) as fixed subalgebras of simply-laced \( \widehat{g} \)
by an outer automorphism \( \widehat{\tau } \). One notes that the exponents
of \( \widehat{g}_{\tau } \) are a subset of the exponents of \( \widehat{g} \)
and it was proven\cite{OTU2} that for this subset of exponents,
\begin{equation}
\label{2.3}
\widehat{\tau }(\widehat{E}_{M})=\widehat{E}_{M\, .}
\end{equation}
It was also shown that 
\begin{equation}
\label{2.4}
\widehat{\tau }(\widehat{F}^{i}_{N})=\widehat{F}^{\tau (i)}_{N}\, .
\end{equation}
 Therefore, the generators of \( \widehat{g} \) which belong to the fixed subalgebra
\( \widehat{g}_{\tau } \) are

\begin{equation}
\label{2.5}
\left\{ \begin{array}{cc}
\widehat{E}_{M} & \mbox {if\, M\, is\, an\, exponent\, of}\, \, \, \widehat{g}_{\tau }\\
\widehat{F}^{<i>}_{N}:=\sum _{i\in <i>}\widehat{F}^{i}_{N} & N\in Z
\end{array}\right. \, .
\end{equation}
From (\ref{2.3}), (\ref{2.4}) and (\ref{1.5}) it follows that
\begin{equation}
\label{2.6}
\gamma _{\tau (i)}\cdot q([M])=\gamma _{i}\cdot q([M])\, .
\end{equation}
 Using this relation, (\ref{2.10a}) and the fact that all the roots in the
orbit \( <i> \) the have same color, it follows that for the generators (\ref{2.5}),
\begin{equation}
\label{2.7}
\left[ \widehat{E}_{M},\widehat{F}^{<i>}_{N}\right] =\gamma _{<i>}\cdot q([M])\widehat{F}_{M+N}^{<i>}\, .
\end{equation}
 Then,

\begin{equation}
\label{2.8}
\widehat{F}^{<i>}(z):=\sum _{i\in <i>}\widehat{F}^{i}(z)
\end{equation}
are in \( \widehat{g}_{\tau } \) and 
\begin{equation}
\label{2.7a}
\left[ \widehat{E}_{M}\, ,\, \widehat{F}^{<i>}(z)\right] =\gamma _{<i>}\cdot q([M])z^{M}\widehat{F}^{<i>}(z)\, ,
\end{equation}
for \( \widehat{E}_{M}\, \in \, \widehat{g}_{\tau } \).

As it has been explained in \cite{KO2}, under a folding procedure, the inequivalent
fundamental representations of \( \widehat{g} \) with highest weights \( \Lambda _{j}\, ,\, \Lambda _{\tau (j)}\, ,\, \Lambda _{\tau ^{2}(j)}\, ,\, ... \)
become identified as a single fundamental representation of \( \widehat{g}_{\tau } \)
whose highest weight is denoted by \( \Lambda _{<j>} \) and with level \( m_{<j>}=m_{j}=m_{\tau (j)}=\, ... \)
.

Recall that as \( \widehat{g} \) is simply laced, in the \( \widehat{g} \)
representation with highest weight \( \Lambda _{j} \), the highest non-vanishing
power of \( \widehat{F}^{i}(z) \) has the vertex operator construction (\ref{1.5a}).
This remains true in the \( \Lambda _{<j>} \) representation of \( \widehat{g}_{\tau } \)
as does \( (F^{i}(z))^{m_{<j>}+1}=0 \). Since\cite{OTU2}
\begin{equation}
\label{2.9}
\left[ \widehat{F}^{i}(z),\widehat{F}^{\tau (i)}(z)\right] =0\, ,
\end{equation}
it follows that in the \( \Lambda _{<j>} \) representation,

\[
\left( \widehat{F}^{<i>}(z)\right) ^{p_{i}m_{<j>}+1}=0\, ,\]
 
\begin{equation}
\label{2.10}
\widehat{V}^{<i>}(z)\equiv \frac{\left( \widehat{F}^{<i>}(z)\right) ^{p_{i}m_{<j>}}}{\left( p_{i}m_{<j>}\right) !}=\prod _{i\in <i>}\frac{\left( \widehat{F}^{i}(z)\right) ^{m_{<j>}}}{m_{<j>}!}\, .
\end{equation}
But each factor in the above product can be written as a vertex operator. Performing
the normal ordering (\ref{1.5c}), using (\ref{2.10b}) and (\ref{2.10c}),
it follows that

\begin{equation}
\label{2.11}
\widehat{V}^{<i>}(z)=\left( b_{<i>}\right) ^{m_{j}}e^{-2\pi i\lambda _{<i>}^{\mbox {v}}\cdot \lambda _{<j>}}Y^{<i>}_{-}Y^{<i>}_{+}
\end{equation}
 where

\begin{equation}
\label{2.12}
Y^{<i>}_{\pm }=\exp \left( \sum _{M\in \cal E}\frac{\gamma ^{\mbox {v}}_{<i>}\cdot q([\pm M])z^{\mp M}}{\mp M}\widehat{E}_{\pm M}\right) \, ,
\end{equation}

\begin{equation}
\label{2.13}
b_{<i>}=\prod _{0\leq p<q\leq p_{i}}X_{\tau ^{p}(i)\tau ^{q}(i)}(z,z)\, .
\end{equation}
The sum in (\ref{2.12}) runs over the positive exponents of \( \widehat{g} \).
However, in order for the vertex operator to make sense in the non-simply laced
algebra \( \widehat{g}_{\tau } \), the sum must be over its exponents. Let
us see that this indeed happens. For the cases in which \( \tau  \) is of order
two and \( M \) is not a degenerate exponent, \( \widehat{\tau }(\widehat{E}_{M})=-\widehat{E}_{M} \)
\cite{OTU2}, and therefore

\begin{equation}
\label{2.14}
\gamma _{i}\cdot q([M])=-\gamma _{\tau (i)}\cdot q([M])\, .
\end{equation}
This implies that \( \gamma _{<i>}^{\mbox {v}}.q([M]) \) will vanish if \( \widehat{E}_{M} \)
does not belong to \( \widehat{g}_{\tau } \). The same result can be obtained
by explicit computation for the order three automorphism of \( d_{4}^{(1)} \)
and for the degenerate exponents of \( d^{(1)}_{2n} \) , by using the fact
that \( \gamma _{i}\cdot q([M]) \) are proportional to the components of the
eigenvectors of the Cartan matrix, which are well-known. Therefore, the above
vertex operator construction makes sense in the non-simply laced algebra \( \widehat{g}_{\tau } \),
since the sum is restricted to its exponents. 

With the above result, we conclude that for \emph{any} affine Lie algebra, in
a level \( m_{j} \) representation, the highest non-vanishing power of \( \widehat{F}^{i}(z) \)
has a vertex operator given by (\ref{2.11}), with \( p_{i}=1 \) and \( b_{i}=1 \)
for the simply laced and twisted algebras.

Using the definition (\ref{2.10}), it is straightforward to check that, in
the \( \Lambda _{<j>} \) representation, 
\begin{equation}
\label{2.15}
\left[ \widehat{E}_{M},\widehat{V}^{<i>}(z)\right] =m_{<j>}\gamma ^{\mbox {v}}_{<i>}\cdot q([M])z^{M}\widehat{V}^{<i>}(z)\, .
\end{equation}
The vertex operator (\ref{2.11}) indeed satisfies this commutation relation.
It is interesting to note that \( \widehat{E}_{M}/m_{<j>} \) and \( \widehat{V}^{<i>}(z) \)
in the \( \Lambda _{<j>} \) representation satisfy the same commutation relation
as \( \widehat{E}_{M} \) and \( \widehat{F}^{<i>}(z) \) satisfy for the dual
algebra \( (\widehat{g}_{\tau })^{\mbox {v}} \).

Using (\ref{1.1}), the normal ordering of two vertex operators can be performed,
in the \( \Lambda _{<j>} \) representation, giving

\begin{equation}
\label{2.16a}
\widehat{V}^{<i>}(z_{i})\widehat{V}^{<k>}(z_{k})=X_{<i><k>}(z_{i},z_{k})^{m_{<j>}}:\widehat{V}^{<i>}(z_{i})\widehat{V}^{<k>}(z_{k}):
\end{equation}
where

\begin{equation}
\label{2.17}
X_{<i><k>}(z_{i},z_{k})=\exp \left( -\sum _{N\in \cal E}\frac{\gamma ^{\mbox {v}}_{i}\cdot q([N\})^{*}\gamma _{k}^{\mbox {v}}\cdot q([N])}{N}\left( \frac{z_{k}}{z_{i}}\right) ^{N}\right) 
\end{equation}
 Similarly to the simply laced case\cite{OTU2}, \( X_{<i><k>} \) can be also
written as 
\begin{equation}
\label{2.18}
X_{<i><k>}(z_{i},z_{k})=\prod _{p=1}^{h}\left( 1-\omega ^{p}\frac{z_{k}}{z_{i}}\right) ^{\sigma ^{-p}\left( \gamma ^{\mbox {v}}_{<i>}\right) \cdot \gamma _{<k>}^{\mbox {v}}}\, ,
\end{equation}
 where \( \omega =\exp (2\pi i/h) \). Moreover, from the commutation relation
(\ref{2.7a}) we find that
\begin{eqnarray}
Y^{<i>}_{+}(z_{i})\widehat{F}^{<k>}(z_{k}) & = & \widehat{F}^{<k>}(z_{k})Y^{<i>}_{+}(z_{i})X_{<i><k>}^{1/p_{k}}(z_{i},z_{k})\label{2.18a} \\
\widehat{F}^{<k>}(z_{k})Y^{<i>}_{-}(z_{i}) & = & Y^{<i>}_{-}(z_{i})\widehat{F}^{<k>}(z_{k})X^{1/p_{k}}_{<k><i>}(z_{k},z_{i})\label{2.18b} 
\end{eqnarray}
which will be very useful in the next sections. Using the fact that 
\[
\sum _{p=1}^{h}\sigma ^{p}\gamma ^{\mbox {v}}_{k}=0\]
and
\[
\sum _{p=1}^{h}p\sigma ^{p}\gamma _{k}=h\left( \lambda _{k}+\Lambda _{R}(g)\right) \, ,\]
 we find that \( X_{<i><k>} \)has the symmetry property
\[
X_{<i><k>}(z_{i},z_{k})^{1/p_{k}}=X_{<k><i>}(z_{k},z_{i})^{1/p_{k}}\]

It is very important to note that all results in this section also hold for
simply laced algebras, just remembering that for these algebras the coroots
(coweights) coincide with the roots(weights) and that the \( p_{i} \)'s are
one. For notational simplicity, from now on the brackets \( <...> \) will no
longer be used.

\section{Asymptotic behavior of soliton solutions.}

Having obtained the vertex operator, the asymptotic behavior of the soliton
solutions (\ref{1.6}) can be checked. In order that one is able to put the
one-soliton solution at rest, the variable \( z_{k} \) must take the form\cite{OTU1}
\begin{equation}
\label{2.19a}
z_{k}=ie^{-\theta _{k}}e^{-i\pi \frac{\left( 1+c(k)\right) }{2h}}
\end{equation}
where \( \theta _{k} \) is the rapidity of the \( k^{\mbox {th}} \) soliton
and \( c(k) \) is the color associated to the \( k \) dot of the Dynkin diagram.
Then, \( X_{ik} \) can be expressed as a function of \( \theta _{ik}\equiv \theta _{i}-\theta _{k} \):
\[
X_{ik}(\theta _{ik})=\prod _{p=1}^{h}\left( e^{\theta _{ik}}-e^{\frac{i\pi }{2h}(4p+c(i)-c(k))}\right) ^{\sigma ^{-p}\gamma _{i}^{\mbox {v}}\cdot \gamma ^{\mbox {v}}_{k}}\]
Now, since \( \widehat{E}_{M}|\Lambda _{j}>=0 \) for \( M>0 \), the \emph{expectation
value} for the vertex operator is
\begin{equation}
\label{2.19}
<\Lambda _{j}|\widehat{V}^{i}(z)|\Lambda _{j}>=b_{i}^{m_{j}}e^{-2\pi i\lambda _{i}^{\mbox {v}}\cdot \lambda _{j}}\, .
\end{equation}
Then, the one-soliton solution \emph{created} by the group element\footnote{
No sum is assumed for the repeated indices.
} \( g(t)=exp(A_{i}\widehat{F}^{i}(\theta )) \), \( A_{i}\equiv Q_{i}W_{i} \),
will take the form 
\begin{equation}
\label{2.20}
e^{-\beta \lambda _{j}\cdot \phi }=\frac{1+\cdots +b_{i}^{m_{j}}e^{-2\pi i\lambda ^{\mbox {v}}_{i}\cdot \lambda _{j}}\left( A_{i}\right) ^{m_{j}p_{i}}}{\left[ 1+\cdots +b_{i}\left( A_{i}\right) ^{p_{i}}\right] ^{m_{j}}}\, ,
\end{equation}
where the dots indicate intermediate powers of \( A_{i} \). These intermediate
powers, whose coefficients we have not calculated, do not affect the asymptotic
limits \( x\rightarrow +\infty  \) or \( x\rightarrow -\infty  \), which are
equivalent to \( A_{i}\rightarrow \infty  \) or \( A_{i}\rightarrow 0 \),
respectively. Thus:
\[
e^{-\beta \lambda _{j}\cdot \phi }=\left\{ \begin{array}{cl}
e^{-2\pi i\lambda _{j}\cdot \lambda _{i}^{\mbox {v}}} & x\rightarrow \infty \\
1 & x\rightarrow -\infty 
\end{array}\right. \]
In particular, this shows that, asymptotically, \( \phi  \) does approach one
of the degenerate vacua. We can also conclude that the topological charge for
the soliton \emph{created} by \( \widehat{F}^{i}(\theta ) \) will satisfy
\[
Q_{top}\equiv \phi (+\infty )-\phi (-\infty )=\frac{2\pi i}{\beta }\left( \lambda ^{\mbox {v}}_{i}+\Lambda _{R}(g_{\tau }^{\mbox {v}})\right) \, .\]
Using the normal ordering (\ref{2.16a}), we obtain that the two-soliton solution
created by \( g(t)=\exp {A_{i}\widehat{F}^{i}(z_{i})} \)~\( \exp {A_{k}\widehat{F}^{k}(z_{k})} \),
takes the form
\[
e^{-\beta \lambda _{j}\cdot \phi }=\frac{1+\cdots +(b_{i}b_{k})^{m_{j}}e^{-2\pi i(\lambda ^{\mbox {v}}_{i}+\lambda ^{\mbox {v}}_{k})\cdot \lambda _{j}}\left( A_{i}^{p_{i}}\, A_{k}^{p_{k}}\right) ^{m_{j}}X_{ik}(\theta _{ik})^{m_{j}}}{\left[ 1+\cdots +b_{i}b_{k}\, A_{i}^{p_{i}}\, A_{k}^{p_{k}}X_{ik}(\theta _{ik})\right] ^{m_{j}}}\, ,\]
 with the asymptotic limits \( \exp \left[ -2\pi i(\lambda _{i}^{\mbox {v}}+\lambda _{k}^{\mbox {v}})\cdot \lambda _{j}\right]  \)
and 1, confirming the expected result that the solution interpolates degenerate
vacua, and with the topological charge satisfying
\[
Q_{top}=\frac{2\pi i}{\beta }\left( \lambda ^{\mbox {v}}_{i}+\lambda _{k}^{\mbox {v}}+\Lambda _{R}(g_{\tau }^{\mbox {v}})\right) \, .\]
 This argument is readily extended to more solitons.

\section{Soliton time delays for non-simply laced ATFT.}

In \cite{FJKO}, the time delay or lateral displacement of the soliton trajectories
arising from collisions has been obtained for simply laced ATFT. Now, we are
in position to extend this result to the non-simply laced ATFT.

Consider a solution (\ref{1.6}) with two solitons created by the group element
\begin{equation}
\label{2.23}
g(t)=e^{A_{i(2)}(\theta _{2})\widehat{F}^{i(2)}(\theta _{2})}e^{A_{i(1)}(\theta _{1})\widehat{F}^{i(1)}(\theta _{1})}
\end{equation}
where we consider that \( \theta _{1}>\theta _{2} \). In order to obtain the
time delay, we shall be \emph{tracking} each soliton in time. By tracking a
soliton ``\( i \)'' we mean remain in its vicinity, which is near \( x=x^{0}_{i}+v_{i}t \).
Following this tracking procedure for soliton 1, which is described in detail
in section 4 of \cite{FJKO}, and using (\ref{2.18a}), (\ref{2.18b}), it follows
that in the past \( t\rightarrow \infty  \), in the vicinity of soliton 1,
\begin{equation}
\label{2.24}
e^{-\beta \lambda _{j}\cdot \phi }\rightarrow \frac{<\Lambda _{j}|e^{A_{i(1)}\widehat{F}^{i(1)}(\theta _{1})}|\Lambda _{j}>}{<\Lambda _{0}|e^{A_{i(1)}\widehat{F}^{i(1)}(\theta _{1})}|\Lambda _{0}>^{m_{j}}}
\end{equation}
which corresponds to an one-soliton solution of species \( i(1) \), velocity
\( v_{1} \), initial position \( x^{0}_{1} \) and phase \( \psi _{1} \).
On the other hand, in the future \( t\rightarrow \infty  \), in the vicinity
of soliton 1,
\begin{equation}
\label{2.25}
e^{-\beta \lambda _{j}\cdot \phi }\rightarrow e^{-2\pi i\lambda ^{\mbox {v}}_{i(2)}\cdot \lambda _{j}}\frac{<\Lambda _{j}|e^{X^{1/p_{i(1)}}_{i(1)i(2)}(\theta _{12})A_{i(1)}\widehat{F}^{i(1)}(\theta _{1})}|\Lambda _{j}>}{<\Lambda _{0}|e^{X_{i(1)i(2)}^{1/p_{i(1)}}(\theta _{12})A_{i(1)}\widehat{F}^{i(1)}(\theta _{1})}|\Lambda _{0}>^{m_{j}}}\, .
\end{equation}
Again, we recognize once more an one-soliton of species \( i(1) \), velocity
\( v_{1} \) and phase \( \psi _{1} \). However, the factor \( X_{i(1)i(2)}^{1/p_{i(1)}} \),
which is real and positive, changes the modulus of \( A_{i(1)} \) and hence
\( x^{0}_{1} \) (see (\ref{1.6c})). The effect is 
\begin{equation}
\label{2.26}
\mu _{i(1)}\cosh \left( x-x^{0}_{1}-v_{1}t\right) \rightarrow \mu _{i(1)}\cosh \left( x-x^{0}_{1}-v_{1}t\right) +\frac{1}{p_{i(1)}}\ln X_{i(1)i(1)}(\theta _{12})\, .
\end{equation}
So, soliton 1 regains its original shape after the collision, and the only effect
of the collision is that solution (\ref{2.24}) differs from (\ref{2.25}) by
a translation in space-time. More precisely, the lateral displacement of soliton
1, due to its collision with soliton 2, \( \Delta _{12}x \), satisfies
\begin{equation}
\label{2.27}
E_{1}\Delta _{12}x=-\frac{M_{i(1)}}{p_{i(1)}\mu _{i(1)}}\ln X_{i(1)i(2)}(\theta _{12})=-\frac{2h}{|\beta ^{2}|}\ln X_{i(1)i(2)}(\theta _{12})\, ,
\end{equation}
where \( E_{1}=M_{i(1)}\cosh \theta _{1} \) is the energy of soliton 1 and
(\ref{2.10c}) has been used as well as the mass formula (\ref{1.6d}). The
time delay \( \Delta _{12}t \) of soliton 1 with momentum \( P_{1} \), is
obtained from 
\[
P_{1}\Delta _{12}t=-E_{1}\Delta _{12}x\, .\]

Following \cite{FJKO}, one can repeat the procedure by tracking the slower
soliton 2, from which it follows that
\begin{equation}
\label{2.29}
E_{2}\Delta _{21}x=\frac{2h}{|\beta ^{2}|}\ln X_{i(1)i(2)}(\theta _{12})=-E_{1}\Delta _{12}x\, ,
\end{equation}
as expected, where \( \Delta _{21}x \) is the lateral displacement of soliton
2 due to its collision with soliton 1. Therefore, we can combine both results
as
\begin{equation}
\label{2.30}
P_{i}\Delta _{ik}t=\mbox {sign}(\theta _{i}-\theta _{k})\frac{2h}{|\beta ^{2}|}\ln X_{ik}(\theta _{ik})\, .
\end{equation}

It is not difficult to extend this procedure from the collision of two solitons
to the collision of any number of solitons. Like \cite{FJKO}, it results for
the \( m^{\mbox {th}} \)soliton that
\[
E_{m}\Delta _{m}x=\frac{2h}{|\beta ^{2}|}\left( \sum _{v_{k}>v_{m}}\ln X_{i(m)i(k)}(\theta _{mk})-\sum _{v_{k}<v_{m}}\ln X_{i(m)i(k)}(\theta _{mk})\right) \]
where \( \Delta _{m}x \) is the total spatial displacement for the \( m^{\mbox {th}} \)
soliton. The left summation are the contributions from the solitons faster than
\( m \) and the right summation are the contributions from the slower solitons.

\section{Semiclassical limit for the soliton S-matrix.}

In 1+1 dimensions, the leading term, in semiclassical approximation, of the
2-state transmission S-matrix elements, can be obtained from the time delays,
as was shown in \cite{JW75}. They proved that for a collision of states \( \alpha  \)
and \( \beta  \) with total energy \( E \) in the centre of momentum frame,
the transmission S-matrix element has the form 
\begin{equation}
\label{5.1}
S_{\alpha \beta }=\exp \left( \frac{i\phi _{\alpha \beta }}{\hbar }+O(\hbar )\right) 
\end{equation}
with
\[
\frac{d\phi _{\alpha \beta }}{dE}=\Delta _{\alpha \beta }t\, .\]
Rewriting this derivative in terms of the relative rapidity, \( \theta \equiv |\theta _{\alpha }-\theta _{\beta }| \)
yields
\[
\frac{d\phi _{\alpha \beta }}{d\theta }=\mbox {sgn}(\theta _{\alpha }-\theta _{\beta })P_{\alpha }\Delta _{\alpha \beta }t\, .\]

Consider the collision of two solitons of species ``\( i \)'' and ``\( k \)''
with respective topological charges ``\( a \)'' and ``\( b \)''. Putting (\ref{2.30})
in the last equation, one gets
\begin{equation}
\label{5.3}
\phi _{ia,kb}(\theta )=\phi _{ia,kb}(0)+\frac{2h}{|\beta ^{2}|}\int ^{\theta }_{0}d\eta \ln X_{ik}(\eta )\, .
\end{equation}
Note that since the time delay does not depend on the soliton topological charges,
then only the integration constant \( \phi _{ia,kb}(0) \) may depend on them. 

Let us denote by \( x_{i}^{(\nu )} \) and \( y^{(\nu )}_{i}=\alpha ^{2}_{i}x_{i}^{(\nu )}/2 \)
the components of the left and right eigenvectors of the Cartan matrix of \( g_{\tau } \)
associated to the common eigenvalue \( 4\sin ^{2}(\pi \nu /2h) \) (\( \nu  \)
being a exponent of \( g_{\tau } \)) and satisfying
\begin{equation}
\label{5.4}
x_{i}^{(\nu )}y_{i}^{(\mu )}=\delta _{\mu \nu }\, \, \, \, \, ,\, \, \, \, \, x^{(\nu )}_{i}y^{(\nu )}_{j}=\delta _{ij}\, .
\end{equation}
 The vectors \( x^{(\nu )}_{i} \) satisfy\footnote{
No sum is assumed for the repeated indices
} 
\begin{equation}
\label{5.5}
c_{i}x_{i}^{(\nu )}=x^{(h-\nu )}_{i}
\end{equation}
 and the inner product \( \gamma _{i}^{\mbox {v}}\cdot q(\nu ) \) can be written
in terms of \( x^{(\nu )}_{i} \) as \cite{FO92} 
\begin{equation}
\label{5.6}
\gamma ^{\mbox {v}}_{i}\cdot q(\nu )=-i\sqrt{2h}e^{\frac{i\pi \nu }{2h}\left( 1+c(i)\right) }x_{i}^{(\nu )}\, .
\end{equation}
Using this result, (\ref{2.19a}) and the exponential form (\ref{2.17}) for
the \( X_{ik}(\eta ) \) function, the above integral can be written as
\begin{equation}
\label{5.7}
\phi _{ia,kb}(\theta )=\phi _{ia,kb}(0)+\frac{4h^{2}}{|\beta ^{2}|}\sum _{N\in \cal E}\overline{x}_{i}^{(N)}\overline{x}^{(N)}_{k}\frac{\left( e^{-N\theta }-1\right) }{N^{2}}
\end{equation}
where \( \overline{x}_{i}^{(\nu +nh)}\equiv [-c(i)]^{n}x^{(\nu )}_{i} \). This
result can be substantially improved: using (\ref{5.5}), the \( \theta  \)
dependent term takes the form 
\begin{equation}
\label{5.8}
\sum _{\nu }x^{(\nu )}_{i}x^{(\nu )}_{k}\left[ \frac{e^{-\nu \theta }}{\nu ^{2}}+\sum _{m=1}^{\infty }\left( \frac{e^{-(2mh+\nu )\theta }}{(2mh+\nu )^{2}}+\frac{e^{-(2mh-\nu )\theta }}{(2mh-\nu )^{2}}\right) \right] \, ,
\end{equation}
where \( \nu  \) are the exponents of the underlying Lie algebra. If we put
\( \theta =0 \), we recognize the term inside the brackets as the series expansion
on simple fractions of the \( \sin ^{-2} \) function\cite{Grad}, resulting
that

\begin{equation}
\label{5.9}
\sum _{N>0}\frac{\overline{x}_{i}^{(N)}\overline{x_{k}}^{(N)}}{N^{2}}=\frac{\pi ^{2}}{h^{2}}\sum _{\nu }\frac{x_{i}^{(\nu )}x_{k}^{(\nu )}}{4\sin ^{2}\frac{\pi \nu }{2h}}=\frac{\pi ^{2}}{h^{2}}\lambda ^{\mbox {v}}_{i}\cdot \lambda ^{\mbox {v}}_{k}\, ,
\end{equation}
where the last equality can be checked directly just using the fact that \( x^{(\nu )} \)
are eigenvectors of the Cartan matrix. 

The infinite series (\ref{5.8}) can be written as a finite sum of polylogarithm
functions, \( \mbox {Li}_{m}(y)\equiv \sum _{n=1}^{\infty }y^{n}/n^{m}\, ,\, |y|<1 \),
by using the identity\footnote{
The author thanks D. Olive for pointing out this identity.
}
\[
\frac{1}{h}\sum _{p=1}^{h}e^{i\pi \frac{p\nu }{h}}\mbox {Li}_{m}\left( e^{-\frac{i\pi p}{h}}x\right) =\sum _{n=1}^{\infty }\frac{x^{nh+\nu }}{\left( nh+\nu \right) ^{m}}\, .\]
Then,

\begin{eqnarray}
\phi _{ia,kb}(\theta ) & = & \phi _{ia,kb}(0)-\frac{4\pi ^{2}}{|\beta ^{2}|}\lambda ^{\mbox {v}}_{i}\cdot \lambda ^{\mbox {v}}_{k}+\nonumber \\
 &  & +\frac{4h^{2}}{|\beta ^{2}|}\sum _{\nu }x^{(\nu )}_{i}x^{(\nu )}_{k}\left( \frac{e^{-\nu \theta }}{\nu ^{2}}+\frac{1}{2h}\sum _{p=1}^{h}\cos \frac{\pi p\nu }{2h}\mbox {Li}_{2}\left( e^{-\frac{i\pi p}{h}-2\theta }\right) \right) \, .
\end{eqnarray}

Alternatively, the phase \( \phi _{ia,kb}(\theta ) \) can be written in terms
of the infinite soliton conserved charges. Indeed, the conserved charges for
the one-soliton solution \emph{created} by \( expQ\widehat{F}^{i}(\theta ) \)
are \cite{Fre95}
\[
P_{i}^{\pm N}(\theta _{i})=-\frac{\mu ^{N}}{|\beta ^{2}|}\gamma ^{\mbox {v}}_{i}\cdot q([\pm N])z_{i}^{\pm N}=\mp \frac{\mu ^{N}}{|\beta ^{2}|}\sqrt{2h}\overline{x}_{i}^{(N)}e^{\pm N\theta _{i}}\, ,\, \, \, N>0\, .\]
Then, 
\begin{equation}
\label{5.12}
\phi _{ia,kb}(\theta )=\phi _{ia,kb}(0)-\frac{4\pi ^{2}}{|\beta ^{2}|}\lambda ^{\mbox {v}}_{i}\cdot \lambda ^{\mbox {v}}_{k}-2h|\beta ^{2}|\sum _{N\in \cal E}\frac{P_{i}^{-N}(\theta _{i})P^{N}_{k}(\theta _{k})}{\mu ^{2N}N^{2}}
\end{equation}
for \( \theta _{i}>\theta _{k} \). This shows an intimate relationship between
soliton S-matrix elements and soliton conserved charges of ATFTs. Indeed, as
pointed out in \cite{BCDS90}\cite{KM91}, the phase shift of any purely elastic
exact S-matrix for an integrable theory could be written as a sum over bilinears
on the \emph{quantum} conserved charges. In the next section, the values of
the coupling in which the soliton S-matrix can be purely elastic are obtained.
So for these values of the coupling, one could expect that an expression like
this for the phase shift would be exact, with \( P^{N}_{i} \) being the quantum
conserved charges and making use of the substitution (\ref{6.1b}), given in
the next section. For example, for sine-Gordon, the integral representation
of \emph{exact} phase shift of the soliton-soliton transmission amplitude\cite{KTTW77}
is

\begin{eqnarray*}
\frac{\phi _{ss}(\theta )}{\hbar } & = & -\frac{1}{2i}\int ^{\infty }_{-\infty }\frac{dt}{t}e^{\frac{2i\theta t}{\pi }}\frac{\sinh (1-\gamma )t}{\sinh \gamma t\cosh t}\\
 & = & \frac{1}{2i}\left[ \int ^{\infty }_{-\infty }\frac{dt}{t}e^{\frac{2i\theta t}{\pi }}-\int ^{\infty }_{-\infty }\frac{dt}{t}e^{\frac{2i\theta t}{\pi }}\tanh t\coth \gamma t\right] \, ,
\end{eqnarray*}
where
\begin{equation}
\label{6.1a}
\gamma =\frac{\hbar |\beta ^{2}|}{4\pi }\left[ 1-\frac{\hbar |\beta ^{2}|}{4\pi }\right] ^{-1}\, .
\end{equation}
 Using the residue theorem it can be written as
\[
\frac{\phi _{ss}(\theta )}{\hbar }=\epsilon \frac{\pi }{2}\left( 1-\frac{1}{\gamma }\right) -\epsilon \left( \sum ^{\infty }_{k=1}\frac{2e^{-\epsilon (2k-1)\theta }}{(2k-1)}\cot \frac{\gamma (2k-1)\pi }{2}-\sum ^{\infty }_{k=1}\frac{e^{-\epsilon \frac{2k\theta }{\gamma }}}{k}\tan \frac{k\pi }{\gamma }\right) \]
where \( \epsilon =\mbox {sgn\, Re}\, \theta  \). One clearly sees that for
\( \gamma =1/n \), \( n=2,3,\, ... \), when the exact reflection amplitude
vanishes, the last summation vanishes and the \emph{exact} phase shift takes
a form similar\footnote{
Remembering that the exponents of \( a^{1}_{1} \) are the odd integers.
} to (\ref{5.7}) or (\ref{5.12}) , with the first term of the series expansion
of the cotangent given by the semiclassical term. From this expression of the
phase shift one gets for free the sine Gordon soliton quantum charges.

It is interesting to note that the ATFT particle S-matrix, which is purely elastic
for any value of the coupling, has a phase shift \cite{Oot97} similar to (\ref{5.12})
but with the conserved charges being proportional to the right eingenvector
\( y_{i}^{(\nu )} \)instead of \( x^{(\nu )}_{i} \). Therefore, for the values
of the coupling in which the soliton S-matrix is purely elastic, there is an
intriguing similarity between the \( \widehat{g} \)-ATFT soliton S-matrix and
the \( \widehat{g}^{\mbox {v}} \)-ATFT particle S-matrix. Clearly for each
particle species there is a soliton species with many possible topological charges.

\section{Number of bound states and purely elastic regime.}

From the above expression for the phase shift, it follows that

\begin{eqnarray}
\phi _{ia,\overline{k}\overline{b}}(0)+\phi _{ia,kb}(0)-\phi _{ia,\overline{k}\overline{b}}(\infty )-\phi _{ia,kb}(\infty ) & = & \frac{4\pi ^{2}}{|\beta ^{2}|}\lambda _{i}^{\mbox {v}}\cdot \left( \lambda _{k}^{\mbox {v}}+\lambda _{\overline{k}}^{\mbox {v}}\right) \nonumber \\
 & \equiv  & n_{ik}(\beta ^{2})\pi \hbar \, ,\label{6.1} 
\end{eqnarray}
 where \( \overline{k}\overline{b} \) means the antisoliton of the soliton
of species \( k \) and topological charge \( b \). Following the super-Levinson
theorem \cite{JW75}, the largest integer smaller than \( n_{ik}(\beta ^{2}) \)
gives the number of bound states in both, the direct and the cross channels.
Note that for sine-Gordon (\( g_{\tau }=a^{(1)}_{1} \) ) \( \lambda _{1}=1/\sqrt{2} \),
and we recover the well-known result\footnote{
The fact that we are adopting the convention \( \alpha _{i}^{2}=2 \) for the
longest simple roots results that our expression differ by a factor 1/2 from
the standard expression.
} that the number of bound states is the largest integer smaller than \( n_{11}=4\pi /\hbar |\beta ^{2}| \),
in the semiclassical approximation. In order to obtain the exact result, one
should replace\cite{DHN75} \( \hbar |\beta ^{2}|/4\pi  \) by \( \gamma  \)
given in (\ref{6.1a}). In view of the arguments in \cite{Dor93}, one would
expect that a similar substitution
\begin{equation}
\label{6.1b}
\frac{\hbar |\beta ^{2}|}{4\pi }\rightarrow \gamma \equiv \frac{\hbar |\beta ^{2}|}{4\pi }\left[ \frac{h}{h^{\mbox {v}}}-\frac{\hbar |\beta ^{2}|}{4\pi }\right] ^{-1}\, .
\end{equation}
 in (\ref{6.1}) would give the exact number of bound states.

These bound states should correspond to poles in the exact S-matrix associated
not just to breathers but also for the fusing of soliton solutions, which appear
as poles of \( X_{ik}(\theta ) \) \cite{FJKO} and are governed by Dorey's
fusing rule\cite{Dor91}. Some of the bound states were analysed\cite{Gan95}
for some particular algebras and particular colliding soliton species.

By direct inspection, one can conclude that 
\begin{equation}
\label{5.14}
\lambda ^{\mbox {v}}_{\overline{k}}=-\lambda _{k}^{\mbox {v}}+\Lambda _{R}(g_{\tau })\, .
\end{equation}
 This result is due to the fact that\cite{KO1} 
\[
\lambda _{\overline{k}}^{\mbox {v}}=-\sigma _{0}(\lambda _{k}^{\mbox {v}})\]
where \( \sigma _{0} \) is the very special element of the Weyl group of \( g^{\mbox {v}}_{\tau } \),
which is the longest one and maps the positive Weyl chamber into its negative.
Note that if we consider 
\begin{equation}
\label{5.15}
\hbar |\beta ^{2}|=\frac{4\pi }{n}\, ,
\end{equation}
where \( n \) is a natural number then, as a consequence of (\ref{5.14}),
\( n_{ik} \) are natural numbers for any \( i \) and \( k \) (and if \( \hbar |\beta ^{2}|>4\pi  \),
\( 0>n_{ik}>1 \), all bound states become unbound). These are exactly the values
of the coupling in which the soliton S-matrix for sine-Gordon becomes purely
elastic\cite{KF78} in the semiclassical approximation. Now we shall show that
same is true for the other ATFTs.

In two dimensions, a two-state S-matrix must satisfy the unitarity and crossing
relations
\begin{eqnarray}
S_{\alpha \beta }^{\mu \nu }(\theta ) & = & \left[ S^{-1}(-\theta )\right] _{\alpha \beta }^{\mu \nu }\, ,\label{6.2} \\
S^{\mu \nu }_{\alpha \beta }(i\pi -\theta ) & = & S^{\mu \overline{\beta }}_{\alpha \overline{\nu }}(\theta )\, .\label{6.3} 
\end{eqnarray}
Equivalently, one can combine both relations and obtain 
\begin{equation}
\label{6.4}
S_{\alpha \beta }^{\mu \nu }(i\pi +\theta )=\left[ S^{-1}(\theta )\right] _{\alpha \overline{\nu }}^{\mu \overline{\beta }}\, .
\end{equation}
In the semiclassical approximation, we must use the crossing relation involving
the analytic continuation \( \theta \rightarrow \theta +i\pi  \) rather than
\( \theta \rightarrow i\pi -\theta  \), since the semiclassical approximation
breaks down on the imaginary axis, as was pointed out by Coleman\cite{Col75}.

Considering that the S-matrix is purely elastic, conditions (\ref{6.2}) and
(\ref{6.4}) become
\begin{eqnarray}
S_{\alpha \beta }(\theta )S_{\alpha \beta }(-\theta ) & = & 1\, ,\label{6.2a} \\
S_{\alpha \beta }(i\pi +\theta ) & = & S_{\alpha \overline{\beta }}^{-1}(\theta )\, ,\label{6.3a} 
\end{eqnarray}
where \( S_{\alpha \beta }(\theta )\equiv S_{\alpha \beta }^{\alpha \beta }(\theta ) \).
Taking \( \theta =0 \) in the first relation, one finds that \( S_{\alpha \beta }(0)=\pm 1 \).
Then, using the semiclassical expression (\ref{5.1}), it follows that
\begin{equation}
\label{6.5}
\phi _{\alpha \beta }(0)=N_{\alpha \beta }\pi \hbar 
\end{equation}
where \( N_{\alpha \beta } \) must be an integer number and gives the number
of bound states in the direct channel\cite{JW75}. 

Let us see the constraint the crossing relation imposes on the phase shift of
ATFT. Before doing this, recall the fact that\cite{FO92} \( c_{\overline{k}}=(-1)^{h}c_{k} \)
and\cite{Dor91} \( x_{\overline{k}}^{(\nu )}=(-1)^{\nu +1}x_{k}^{(\nu )} \).
Then, it follows from the definition of \( \overline{x}_{k}^{(N)} \), that

\[
\overline{x}^{(N)}_{k}e^{-i\pi N}=-\overline{x}^{(N)}_{\overline{k}}\, .\]
This relation implies that 
\[
\phi _{ia,kb}(\theta +i\pi )=\phi _{ia,kb}(0)-\frac{4\pi ^{2}}{|\beta ^{2}|}\lambda ^{\mbox {v}}_{i}\cdot \lambda ^{\mbox {v}}_{k}-\frac{4h^{2}}{|\beta ^{2}|}\sum _{N\in \cal E}\frac{\overline{x}_{i}^{(N)}\overline{x}^{(N)}_{\overline{k}}}{N^{2}}e^{-N\theta }\, .\]
Then, the crossing condition (\ref{6.3a}) results that 
\begin{equation}
\label{6.6}
\phi _{ia,\overline{k}\overline{b}}(0)=-\phi _{ia,kb}(0)+\frac{4\pi ^{2}}{|\beta ^{2}|}\lambda _{i}^{\mbox {v}}\cdot \left( \lambda ^{\mbox {v}}_{k}+\lambda _{\overline{k}}^{\mbox {v}}\right) \, \, \, \, \, \, \, (\mbox {mod}2\pi \hbar )
\end{equation}
Using (\ref{5.14}) and the fact that \( \phi _{ia,\overline{k}\overline{b}}(0) \)
and \( \phi _{ia,kb}(0) \) must satisfy (\ref{6.5}), it implies that the above
equation can only be consistent if the coupling satisfies the condition (\ref{5.15}).
Therefore, only for these values of the coupling the unitarity and crossing
relations for purely elastic S-matrix is fulfilled. So, the S-matrix can only
be purely elastic for these values in the semiclassical approximation. Once
more, this result holds at the semiclassical approximation and in order to obtain
the exact result one should probably perform a substitution like (\ref{6.1b}).
Comparing relation (\ref{6.6}) with the super-Levinson relation (\ref{6.1})
implies that \( \phi _{ia,\overline{k}\overline{b}}(\infty )+\phi _{ia,kb}(\infty )=0 \)
mod\( (2\pi \hbar ) \). It would be interesting if one could calculate the
constants \( \phi _{ia,kb}(0) \), which gives the number of bound states in
the direct channel\cite{JW75}.

\section{Conclusions.}

In this paper, we have extended our vertex operator construction for non-simply
laced affine Lie algebras from which we obtained the time delay resulting from
the collision of two (or more) solitons. From the time delay, we obtained the
semiclassical limit for the transmission amplitude of the two soliton S-matrix
which holds for any ATFT. The semiclassical phase shift appeared as a sum over
bilinears on the soliton conserved charges. Using the super-Levison theorem,
an universal expression for the number of bound states in the direct and cross
channels of the S-matrix was obtained. We also obtained the values of the coupling
in which the S-matrix can be purely elastic. For these values, one would expect
that the form of the exact phase shift would be like (\ref{5.12}), with \( P^{i}_{N} \)
being the \emph{quantum} soliton conserved charges. 

\vskip 0.2 in \noindent \textbf{Acknowledgments }

\noindent I would like to thank D. Olive and R. Paunov for discussions and for
reading the manuscript, L.A. Ferreira for invitation to visit IFT-UNESP where
part of this work was conceived and FAPERJ for financial support.

\end{document}